\documentclass[prd,nopacs,nokeys]{revtex4}
\usepackage{amsfonts}
\usepackage{amsmath}
\usepackage{amssymb}
\usepackage{amsmath}
\usepackage{bbm}
\usepackage[utf8]{inputenc}
\usepackage[caption=false]{subfig}
\usepackage{tensor}
\usepackage{slashed}
\usepackage[centertableaux ]{ytableau}
\usepackage{hyperref}
\usepackage{natbib}
\usepackage{braket,mleftright}
\usepackage{graphicx}
\usepackage{cleveref}
\usepackage{tikz}
\usepackage{tikz,ifthen}

\usetikzlibrary{trees}
\usetikzlibrary{decorations.pathmorphing}
\usetikzlibrary{decorations.markings}
\tikzset{beamerprimary/.style={structure.fg, thick}}
\tikzset{beamersecondary/.style={structure.bg, thick}}
\tikzset{boson/.style={draw=structure.fg,decorate, decoration={snake}},
    gauge/.style={decorate, decoration={snake} },
    fermion/.style={postaction={decorate},
         decoration={markings,mark=at position .55 with {\arrow{>}}}},
    fermionloop/.style={postaction={decorate},
        },
    gluon/.style={decorate,
        decoration={coil,amplitude=4pt, segment length=5pt}},
    scalar/.style={dashed},
    graviton/.style={double}
}

\begin{document}

\title{Effects of an $H-\mu-\tau $ coupling in quarkonium lepton flavor
violation decays}
\author{David Delepine and Mauro Napsuciale}
\affiliation{Universidad de Guanajuato\\
Lomas del Bosque 103, Fraccionamiento Lomas del Campestre, 37150, Le\'on,
Guanajuato, M\'{e}xico}
\author{Eduardo Peinado}
\affiliation{Instituto de F\'{i}sica, Universidad Nacional Aut\'onoma de M\'{e}xico \\
A.P. 20-364, M\'exico D.F. 01000, M\'exico.}

\begin{abstract}
In this work we study the consistency of a possible non-vanishing coupling 
$H\mu \tau $ of the order of $3.6\times 10^{-3}$ as pointed recently by the
CMS and ATLAS collaborations \cite{Khachatryan:2015kon}, \cite{Aad:2015gha}, 
with measured lepton flavor violation processes involving quarkonium. We show 
that the most promising channel to confirm this excess is to look for the lepton 
flavor tau decay into a $f_0$ and $\mu$ where the experimental limit could 
strongly improved with the new B factories as Belle II. 
\end{abstract}

\maketitle

%\section{Introduction}

Recently CMS collaboration has observed  a slight excess of signal events with 
a significance of 2.4 standard deviations which can be interpreted as a Higgs 
particles decaying into a muon and tau leptons:
\begin{equation}
Br(H \rightarrow \mu \tau)=\left\{
\begin{tabular}{c}
$ 0.84^{+0.39}_{-0.37} ~\% $ \, \cite{Khachatryan:2015kon} \\
$0.77\pm 0.62 ~\% $ \cite{Aad:2015gha} \\
\end{tabular}
\right.
\end{equation}
Using the reported value of Higgs mass to be $125$ GeV~\cite{Aad:2012tfa}, 
this requires the coupling $H\mu \tau $ to be of the order of $3.6\times 10^{-3}$. 
Even if this observation is very challenging to be explained in new physics 
models~\cite{NP}, this value of the lepton flavor violating coupling is not in 
contradiction with the experimental upper limits for $\tau \to \mu\gamma$ and $\tau\to 3\mu$. 
Indeed, using the formalism in Refs.~\cite{McWilliams:1980kj,Harnik:2012pb} we obtain the 
values listed in Table \ref{table11} for the corresponding branching ratios. 
\begin{table}
\begin{tabular}{|c|c|c|}
  \hline
   & Experimental bound \cite{Agashe:2014kda} & Expected from $H \rightarrow \mu \tau$ \\
   \hline
  $\tau \rightarrow \mu \gamma$ & $4.4 \times 10^{-8}$  & $1.3 \times 10^{-9}  $ \\
  $\tau \rightarrow 3 \mu$ & $2.1 \times 10^{-8}$ & $1\times 10^{-10}  $\\
  \hline
\end{tabular}
\caption{LFV in $\tau$ LFV decays involving charged leptons}
\label{table11}
\end{table}
The expected value from $H \rightarrow \mu \tau$ is done assuming that the Higgs 
couplings to charged leptons are given by SM values\cite{Chatrchyan:2014vua,Chatrchyan:2014nva}.
 
 The computation of Higgs-induced lepton flavor violation (LFV) in channel decays 
 involving only charged leptons are even more intricated as the smallness of the 
 lepton Yukawa couplings imply that higher loop contributions can be bigger than 
 tree-level one \cite{Goudelis:2011un,Harnik:2012pb}. To avoid this problem, we 
 shall study the effect of the LFV $H \rightarrow \tau \mu$ coupling in  processes 
 involving quarkonium. In Ref.~\cite{Abada:2015zea} the effects of heavy sterile 
 Majorana neutrinos in LFV decays of vector quarkonia has been studied. Sterile 
 Majorana neutrinos induces $\gamma l_i l_j$, $Zl_i l_j$ LFV couplings at one loop and 
 $WW l_i l_j$ LFV couplings at tree level. These couplings produce LFV effects in quarkonia 
 decay studied in Ref. \cite{Abada:2015zea}. Another effect not studied there, is to produce 
 a non-vanishing $H l_i l_j$ LFV coupling at one level, whose effects in quarkonium decay 
 were no analyzed in Ref. \cite{Abada:2015zea}. 
 
 In this paper, we assume no specific models for new physics behind the $H \tau \mu$ coupling 
 and we systematically study the LFV decays of quarkonia involving the $H \tau \mu$ coupling, 
 considering the phenomenological value of $3.6\times 10^{-3}$ pointed by the CMS and  ATLAS collaborations.  We shall show that even if the expected branching ratio are 
 still below the experimental limit, some of them could be accessible to next generation 
 of $B$ factories as Belle II.

The description of processes involving the annihilation or creation  of heavy quarkonia
can be systematically done in the framework of non-relativistic quantum chromodynamics 
(NRQCD) \cite{Bodwin:1994jh}. This is a systematic expansion in terms of $\alpha_s$ and the quarks 
relative velocity $v$ with a clear separation of the perturbative phenomena occurring at the 
scale $m_{Q}$ and the non-perturbative ones occurring at the scale $m_{Q}v$. The 
non-perturbative effects are encoded in universal matrix elements 
with a well defined hierarchy in the the $v$ expansion. The novelty of this 
systematic approach is that, for some processes, color-octet configurations of the created or annihilated 
quark-antiquark pair  yield contributions of the same order as the old color-singlet contributions 
to a given order in the $\alpha_s$ and $v$ expansion. 

In this work we are interested in the order of magnitude of the 
branching ratios of the considered processes and will focus on the color singlet contributions 
which can be calculated using the old quarkonium techniques described in 
\cite{Kuhn:1979bb, Guberina:1980dc}. A more refined analysis can be done in the most 
promising channels but this is beyond the scope of the present work.
  
The invariant amplitude for the annihilation of color-singlet quarkonium in a $^{2S+1}L_{J}$ angular 
momentum configuration $\overline{Q}Q[^{2S+1}L_{J}]\rightarrow X$ is given by 
\cite{Kuhn:1979bb, Guberina:1980dc}%
\begin{equation}
\mathcal{M}[\overline{Q}Q[^{2S+1}L_{J}]\rightarrow X]=\int \frac{d^{4}q}{%
  (2\pi )^{4}}Tr[\mathcal{O}(Q,q)\chi (Q,q)],  \label{maineq}
\end{equation}%
where $\mathcal{O}(Q,q)$ is the operator entering amplitude for the
corresponding free quarks transition
\begin{equation}
\mathcal{M}[\overline{Q}(\frac{Q}{2}-q,s_{2}),Q(\frac{Q}{2}%
+q,s_{1})\rightarrow X]=\overline{v}(\frac{Q}{2}-q,s_{2})\mathcal{O}(Q,q)u(%
\frac{Q}{2}+q,s_{1}),
\end{equation}%
and $\chi (Q,q)$ denotes the wave function for the $\overline{Q}%
Q[^{2S+1}L_{J}]$ bound state%
\begin{equation}
\chi (Q,q)=\sum\limits_{M,S_{z}}2\pi \delta (q^{0}-\frac{\mathbf{q}^{2}}{%
2m_{Q}})\psi _{LM}(\mathbf{q})P_{S,S_{z}}(Q,q)\langle
LM;SS_{z}|JJ_{z}\rangle .
\end{equation}%
Here, $P_{S,S_{z}}$ stands for the spin projectors
\begin{eqnarray}
P_{S,S_{z}}(Q,q) &=&\sqrt{\frac{N_{c}}{m_{Q}}}\sum\limits_{s_{1},s_{2}}u(%
\frac{Q}{2}+q,s_{1})\overline{v}(\frac{Q}{2}-q,s_{2})\langle \frac{1}{2}%
s_{1};\frac{1}{2}s_{2}|SS_{z}\rangle \\
&=&\sqrt{\frac{N_{c}}{m_{Q}}}\left( \frac{1}{2\sqrt{2}m_{Q}}\right) (\frac{%
\slashed{Q}}{2}+\slashed{q}+m_{Q})\left\{
\begin{tabular}{l}
$\gamma ^{5}$ \\
$\slashed{\varepsilon}(Q,S_{z})$%
\end{tabular}%
\right\} (\frac{\slashed{Q}}{2}+\slashed{q}-m_{Q})\text{ for }\left\{
\begin{tabular}{l}
$S=0$ \\
$S=1$%
\end{tabular}%
\right\} ,
\end{eqnarray}%
where $\varepsilon (Q,S_{z})$ denotes the polarization vector of the spin one system.

For $s$-wave quarkonium the wave function is rapidly damped in the relative
momentum $q$ and the leading terms are given by $P_{S,S_{z}}(Q,0)$ and $\mathcal{O}(Q,0)$. 
In the zero-binding approximation
the quarkonium mass $M$ is given by $M\approx 2m_{Q}$ and the amplitude
reads
\begin{equation}
\mathcal{M}[\overline{Q}Q[^{2S+1}S_{J}]\rightarrow X]=\frac{R(0)}{\sqrt{4\pi
}}\sqrt{\frac{3}{4M}}Tr\left[ \mathcal{O}(Q,0)\left\{
\begin{tabular}{l}
$\gamma ^{5}$ \\
$\slashed{\varepsilon}(Q,S_{z})$%
\end{tabular}%
\right\} (\slashed{Q}-M)\right] \text{for }\left\{
\begin{tabular}{l}
$S=0$ \\
$S=1$%
\end{tabular}%
\right\},  \label{swave}
\end{equation}%
with $M$ denoting the quarkonium physical mass and%
\begin{equation}
\int \frac{d^{3}q}{(2\pi )^{3}}\psi _{00}(\mathbf{q})=\frac{R(0)}{\sqrt{4\pi
}}.
\end{equation}%
A similar calculation of the invariant amplitude for the production of color
singlet quarkonium, $X\rightarrow $ $\overline{Q}Q[^{2S+1}S_{J}]+Y$ yields%
\begin{equation}
\mathcal{M}(X\rightarrow  \overline{Q}Q[^{2S+1}S_{J}]+Y)=-\frac{R(0)}{\sqrt{%
4\pi }}\sqrt{\frac{3}{4M}}Tr\left[ \mathcal{O}(Q,0)\left\{
\begin{tabular}{l}
$\gamma ^{5}$ \\
$\slashed{\varepsilon}(Q,S_{z})$%
\end{tabular}%
\right\} (\slashed{Q}+M)\right] \text{for }\left\{
\begin{tabular}{l}
$S=0$ \\
$S=1$%
\end{tabular}%
\right\} .
\end{equation}

For $p$-wave quarkonium, the wave function at the origin vanishes and the
leading term for the annihilation amplitude is given by the first term in
the expansion in $q$ of Eq.~(\ref{maineq}). A straightforward calculation yields%
\begin{equation}
\mathcal{M}[\overline{Q}Q[^{2S+1}P_{J}]\rightarrow
X]=-i\sum\limits_{M,S_{z}}\langle 1M;SS_{z}|JJ_{z}\rangle \varepsilon
_{\alpha }(M)\sqrt{\frac{3}{4\pi }}R^{\prime }(0)Tr\left[ \mathcal{O}%
^{\alpha }(Q,0)P_{S,S_{z}}(Q,0)+\mathcal{O}(Q,0)P_{S,S_{z}}^{\alpha }(Q,0)%
\right] ,
\end{equation}%
where%
\begin{equation}
A^{\alpha }(Q,q)\equiv \frac{\partial A(Q,q)}{\partial q_{\alpha }},
\end{equation}%
and in this case%
\begin{equation}
\int \frac{d^{3}q}{(2\pi )^{3}}q^{\alpha }\psi _{1M}(\mathbf{q})=-i\sqrt{%
\frac{3}{4\pi }}R^{\prime }(0)\varepsilon _{\alpha }(M).
\end{equation}%
The polarization vector $\varepsilon _{\alpha }(M)$ satisfies the following
relations%
\begin{eqnarray}
\sum\limits_{M,S_{z}}\langle 1M;1S_{z}|00\rangle \varepsilon _{\alpha
}(M)\varepsilon _{\beta }(S_{z}) &=&-g_{\alpha \beta }+\frac{Q_{\alpha
}Q_{\beta }}{M^{2}}, \\
\sum\limits_{M,S_{z}}\langle 1M;1S_{z}|1J_{z}\rangle \varepsilon _{\alpha
}(M)\varepsilon _{\beta }(S_{z}) &=&\frac{-i}{M}\frac{1}{\sqrt{2}}%
\varepsilon _{\alpha \beta \mu \nu }Q^{\mu }\varepsilon ^{\nu }(J_{z}), \\
\sum\limits_{M,S_{z}}\langle 1M;1S_{z}|2J_{z}\rangle \varepsilon _{\alpha
}(M)\varepsilon _{\beta }(S_{z}) &=&\varepsilon _{\alpha \beta }(J_{z}).
\end{eqnarray}

%\section{Flavor changing transitions involving quarkoniuum}

%\subsection{Higgs-quarkonium coupling}

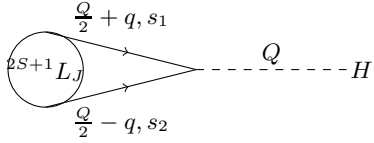
\begin{figure}[ht]
\centering
\begin{tikzpicture}
\draw[fermion] (-2,0.5) --(0,0) ;
\draw[fermion] (-2,-0.5)--(0,0) ;
\draw[fermionloop] (-2.0,0) circle (.5);
\draw[scalar] (0,0)--(2,0) ;
\node at (-2,0) {$^{2S+1}L_{J}$};
\node at (2.2,0) {$H$};
\node at (1,0.2) {$Q$};
\node at(-1,0.7){$\frac{Q}{2}+q,s_{1}$};
\node at(-1,-0.7){$\frac{Q}{2}-q,s_{2}$};
\end{tikzpicture}
\caption{Feynman diagrams for the Higgs-quarkonium coupling.}
\label{HQQ}
\end{figure}

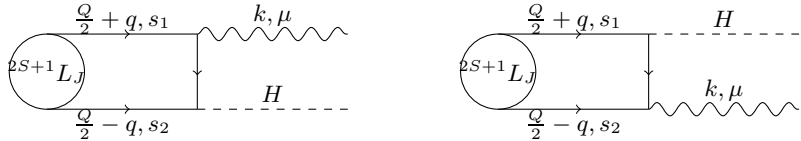
\begin{figure}[ht]
\centering
\begin{tikzpicture}
%First diagram
\draw[fermion] (-5,0.5) --(-3,0.5) ;
\draw[gauge] (-3,0.5)  -- node[above]{$k,\mu$}(-1,0.5);
\draw[fermion] (-3,0.5) --(-3,-0.5) ;
\draw[fermion] (-5,-0.5)--(-3,-0.5) ;
\draw[fermionloop] (-5.0,0) circle (.5);
\draw[scalar] (-3,-0.5)--(-1,-0.5) ;
\node at (-5,0) {$^{2S+1}L_{J}$};
\node at (-2,-0.3) {$H$};
\node at(-4,0.7){$\frac{Q}{2}+q,s_{1}$};
\node at(-4,-0.7){$\frac{Q}{2}-q,s_{2}$};
%Second diagram
\draw[fermion] (1,0.5) --(3,0.5) ;
\draw[gauge] (3,-0.5)  -- node[above]{$k,\mu$}(5,-0.5);
\draw[fermion] (3,0.5) --(3,-0.5) ;
\draw[fermion] (1,-0.5)--(3,-0.5) ;
\draw[fermionloop] (1.0,0) circle (.5);
\draw[scalar] (3,0.5)--(5,0.5) ;
\node at (1.0,0) {$^{2S+1}L_{J}$};
\node at (4,0.7) {$H$};
\node at(2,0.7){$\frac{Q}{2}+q,s_{1}$};
\node at(2,-0.7){$\frac{Q}{2}-q,s_{2}$};
\end{tikzpicture}
\caption{Feynman diagrams for the Higgs-quarkonium-photon coupling.}
\label{HQQg}
\end{figure}

The Higgs to $\overline{Q}Q[^{2S+1}L_{J}]$ quarkonium coupling is obtained
from the diagram in Fig. (\ref{HQQ}) which yields the following operator
\begin{equation}
\mathcal{O}(Q,q)=i\frac{m_{Q}}{v},
\end{equation}%
where $v$ stands for the Higgs vacuum expectation value.
%
%%%%%%
%%%%%
Using this operator in the previous formulae it is easy to show that the only 
non-vanishing coupling of the Higgs to quarkonium is to $S=1$, $J=0$ 
$p$-wave quarkonium, in which case we obtain%
\begin{equation}
\mathcal{M}[\overline{Q}Q[^{3}P_{0}]\rightarrow H]=
\frac{3R^{\prime }(0)}{v}\sqrt{\frac{3M}{\pi }}.
\end{equation}
Notice that this coupling is proportional to the derivative of the wave
function at the origin, which according to the NRQCD rules is suppressed by
a $v^2$ factor with respect to the wave function at the
origin. This makes the radiative transitions involving $s$-wave quarkonium
configurations of the same order as the non-radiative ones involving $p$-wave 
quarkonium configurations. The radiation changes the quarkonium quantum numbers 
allowing the corresponding quarkonium to couple to the Higgs. 
The calculation of Higgs-mediated lepton flavor
violating radiative transitions involving $s$-wave quarkonium requires to work out Higgs-quarkonium-photon coupling.
%\subsection{$H-\overline{Q}Q[^{2S+1}S_{J}]-\protect\gamma$ coupling}
This transition is induced by the diagrams in Fig.~(\ref{HQQg}).
%%%%%%
%%%%%

From these diagrams, we identify the transition operator as
\begin{equation}
\mathcal{O}(Q,q)=iee_{Q}\frac{m_{Q}}{v} \left[\slashed{\varepsilon}(k)%
\frac{\frac{\slashed{Q}}{2}+\slashed{q}+\slashed{k}+m_{Q} } {(\frac{Q}{2}%
+q+k)^{2}-m^{2}_{Q}} - \frac{\frac{\slashed{Q}}{2}-\slashed{q}+\slashed{k}%
-m_{Q} } {(\frac{Q}{2}-q+k)^{2}-m^{2}_{Q}}\slashed{\varepsilon}(k) \right].
\end{equation}%
where $e_{Q}$ stands for the heavy quark charge in units of $e$. For $s$%
-wave $J=0$ from Eq.~(\ref{swave}) we obtain
\begin{equation}
\mathcal{M}[H\rightarrow \overline{Q}Q[^{1}S_{0}]\gamma]=
-i\frac{e e_{Q}M R(0)}{4v}\sqrt{\frac{3}{4\pi M}}Tr\left[ \left(\frac{%
\slashed{\varepsilon}(k)(\frac{\slashed{Q}}{2}+\slashed{k}) -(\frac{%
\slashed{Q}}{2}+\slashed{k}) \slashed{\varepsilon}(k)} {(\frac{Q}{2}%
+k)^{2}-m^{2}_{Q}} \right) \gamma ^{5} (\slashed{Q}-M)\right]=0.
\end{equation}
Similarly for $s$-wave $J=1$ we get
\begin{equation}
\mathcal{M}[H\rightarrow \overline{Q}Q[^{3}S_{1}]\gamma]=i\frac{%
ee_{Q}m_{Q}R(0)}{2 v}\sqrt{\frac{3}{4\pi M}}Tr\left[ \left(\frac{%
\slashed{\varepsilon}(k)(\frac{\slashed{Q}}{2}+\slashed{k}) -(\frac{%
\slashed{Q}}{2}+\slashed{k}) \slashed{\varepsilon}(k)} {(\frac{Q}{2}%
+k)^{2}-m^{2}_{Q}} \right) \slashed{\eta}(Q) (\slashed{Q}-M)\right],
\end{equation}
where $\eta(Q)$ stands for the polarization vector of the quarkonium. A
straightforward calculation yields
\begin{equation}
\mathcal{M}[H\rightarrow \overline{Q}Q[^{3}S_{1}]\gamma]=\frac{ee_{Q}R(0)}{%
v}\sqrt{\frac{3M}{\pi}} T_{\mu\nu}\varepsilon^{\mu}\eta^{\nu}
\end{equation}
with
\begin{equation}
T_{\mu\nu}=g_{\mu\nu}-\frac{Q^{\mu}k^{\nu}}{Q \cdot k}.
\end{equation}

\begin{figure}[ht]
\centering
\begin{tikzpicture}
\draw[fermion] (-2,0.5) --(0,0) ;
\draw[fermion] (-2,-0.5)--(0,0) ;
\draw[fermionloop] (-2.0,0) circle (.5);
\draw[scalar] (0,0)--(2,0) ;
\node at (-2,0) {$^{3}P_{0}$};
\node at (1,0.2) {$H$};
\node at(-1,0.7){$\frac{Q}{2}+q,s_{1}$};
\node at(-1,-0.7){$\frac{Q}{2}-q,s_{2}$};
\draw[fermion] (2,0)--node[above]{$\mu$}(3,0.5) ;
\draw[fermion] (2,0)--node[below]{$\tau$}(3,-0.5) ;
\end{tikzpicture}
\caption{Diagram for the $\overline{Q}Q[^{3}P_{0}]\to \protect\mu\protect%
\tau $ decay.}
\label{chitomutau}
\end{figure}
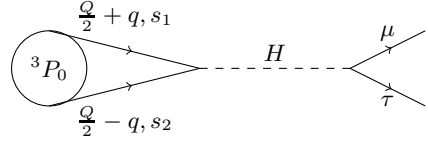

%\subsection{Flavor violating Quarkonium decays}
%\subsubsection{$\overline{Q}Q[^{3}P_{0}]\to \protect\mu\protect\tau$}
Now we focus on the Higgs mediated LFV processes. We start with 
the $\overline{Q}Q[^{3}P_{0}]\to \protect\mu\protect\tau$ decay through the 
diagram in Fig (\ref{chitomutau}).
%%%%%%
%%%%%
A direct calculation yields the following decay width
\begin{equation}
\Gamma[{{\overline{Q}Q[^{3}P_{0}]\to \mu^{-}\tau^{+}}}]=\frac{27 y^{2}}{8\pi^2}
\frac{|R^{\prime}(0)|^{2}M^{2}}{v^{2}m^{4}_H} \left(1-\frac{m^{2}_{\tau}}{M^{2}}%
\right)
\end{equation}
where we neglected the muon mass and $y$ stands for the $H\mu\tau$ coupling.

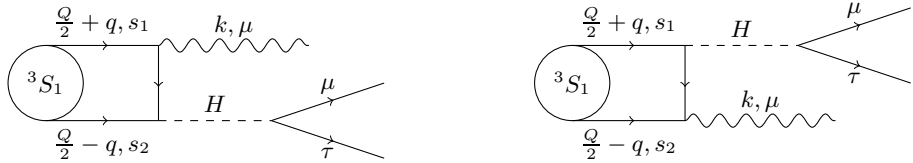
\begin{figure}[ht]
%\centering
\begin{tikzpicture}
%First diagram
\draw[fermion] (-6,0.5) --node[above]{$\frac{Q}{2}+q,s_{1}$}(-4.5,0.5) ;
\draw[gauge] (-4.5,0.5)  -- node[above]{$k,\mu$}(-2.5,0.5);
\draw[fermion] (-4.5,0.5) --(-4.5,-0.5) ;
\draw[fermion] (-6,-0.5)--node[below]{$\frac{Q}{2}-q,s_{2}$}(-4.5,-0.5) ;
\draw[fermionloop] (-6.0,0) circle (.5);
\draw[scalar] (-4.5,-0.5)--node[above]{$H$}(-3,-0.5) ;
\draw[fermion] (-3,-0.5)--node[above]{$\mu$}(-1.5,0) ;
\draw[fermion] (-3,-0.5)--node[below]{$\tau$}(-1.5,-1) ;
\node at (-6,0) {$^{3}S_{1}$};
%Second diagram
\draw[fermion] (1,0.5) --node[above]{$\frac{Q}{2}+q,s_{1}$}(2.5,0.5) ;
\draw[gauge] (2.5,-0.5)  -- node[above]{$k,\mu$}(4.5,-0.5);
\draw[fermion] (2.5,0.5) --(2.5,-0.5) ;
\draw[fermion] (1,-0.5)--node[below]{$\frac{Q}{2}-q,s_{2}$}(2.5,-0.5) ;
\draw[fermionloop] (1.0,0) circle (.5);
\draw[scalar] (2.5,0.5)--node[above]{$H$}(4,0.5) ;
\draw[fermion] (4,0.5)--node[above]{$\mu$}(5.5,1) ;
\draw[fermion] (4,0.5)--node[below]{$\tau$}(5.5,0) ;
\node at (1,0) {$^{3}S_{1}$};
\end{tikzpicture}
\caption{Feynman diagrams for the $^{3}S_{1}\to \protect\mu\protect\tau$
decay.}
\label{3S1tomutau}
\end{figure}

%\subsubsection{$\overline{Q}Q[^{3}S_{1}]\to \protect\mu\protect\tau \protect\gamma$}
Next we go through the corresponding radiative process 
$\overline{Q}Q[^{3}S_{1}]\to \protect\mu\protect\tau \protect\gamma$. This decay proceeds through the diagrams in Fig. (\ref{3S1tomutau}). 
%%%%%%
%%%%%

Neglecting the muon mass we obtain the following decay width
\begin{equation}
\Gamma[{{\overline{Q}Q[^{3}S_{1}]\to \mu^{-}\tau^{+}\gamma}}]=\frac{\alpha e^2_{Q}
y^{2} |R(0)|^{2}}{12\pi^3v^{2}} \frac{M^4}{m^{4}_H} f\left(\frac{m^{2}_{\tau}}{M^2}\right),
\end{equation}
where
\begin{equation}
f(x)=1-6x+3x^2+2x^3-6~ \ln(x).
\end{equation}

%\subsection{Flavor violating $\protect\tau$ decays to quarkonium}
%\subsubsection{$\protect\tau \rightarrow \protect\mu ~\overline{Q}Q[^{3}P_{0}]$}
The $H\mu\tau$ coupling can also mediate LFV decays of the tau meson involving light 
quarkonium. Although this is beyond the scope of the systematic NRQCD expansion due to 
the light quark mass, we still can use the quarkonium techniques taking care of extracting the 
corresponding non-perturbative pieces from the appropriate  experimental data.
The first possible decay is $\protect\tau \rightarrow \protect\mu ~\overline{Q}Q[^{3}P_{0}]$ 
which goes through the diagram shown in Fig.~(\ref{tautomuf}).
%%%%%%
\begin{figure}[ht]
\centering
\begin{tikzpicture}
\draw[fermion] (-3,0)--node[above]{$\tau$}(-1,0) ;
\draw[fermion] (-1,0)--node[above]{$\mu$}(2,-1) ;
\draw[scalar] (-1,0)--(1,0.5) ;
\draw[fermion] (1,0.5) --(3,1) ;
\draw[fermion] (1,0.5)--(3,0) ;
\draw[fermionloop] (3.0,0.5) circle (.5);
\node at (3,0.5) {$^{3}P_{0}$};
\node at (0,0.5) {$H$};
\node at(2,1.1){$\frac{Q}{2}+q,s_{1}$};
\node at(2,0){$\frac{Q}{2}-q,s_{2}$};
\end{tikzpicture}
\caption{Feynman diagram for the $\protect\tau\to \protect\mu \overline{Q}%
Q[^{3}P_{0}]$ decay.}
\label{tautomuf}
\end{figure}
%%%%%
The decay width is given by%
\begin{equation}
\Gamma (\tau^{-} \rightarrow \mu^{-} ~\overline{Q}Q[^{3}P_{0}])=\frac{%
27y^{2}|R^{\prime}(0)|^{2}}{16\pi ^{2}v^2}\frac{m_{\tau }M}{m_{H}^{4}}\left(1-\frac{%
M^{2}}{m_{\tau }^{2}}\right),
\end{equation}%
where we neglected the muon mass.

%\subsubsection{$\protect\tau \rightarrow \protect\mu ~\overline{Q}Q[^{3}S_{1}]\protect\gamma $}
The corresponding radiative decay is 
$\protect\tau \rightarrow \protect\mu ~\overline{Q}Q[^{3}S_{1}]\protect\gamma $.
The Feynman diagrams for this process are given in Fig. (\ref{tautomu3S1gamma}). 
%%%%%%
\begin{figure}[ht]
%\centering
\begin{tikzpicture}
%First diagram
\draw[fermion] (-5.5,0)--node[above]{$\tau$}(-4,0) ;
\draw[fermion] (-4,0)--node[above]{$\mu$}(-2,-1) ;
\draw[scalar] (-4,0)--node[above]{$H$}(-2.5,0.5) ;
\draw[fermion] (-2.5,0.5) --(-1.5,0.5) ;
\draw[fermion] (-1.5,0.5) --(-0.5,0.5) ;
\draw[fermion] (-2.5,0.5)--(-0.5,-0.5) ;
\draw[fermionloop] (-0.5,0) circle (.5);
\draw[gauge] (-1.5,0.5)  -- (0,1);
\node at (-0.5,0) {$^{3}S_{1}$};
%Second diagram
\draw[fermion] (0.5,0)--node[above]{$\tau$}(2,0) ;
\draw[fermion] (2,0)--node[above]{$\mu$}(4,-1) ;
\draw[scalar] (2,0)--node[above]{$H$}(3,0.5) ;
\draw[fermion] (3,0.5) --(4,0.5) ;
\draw[fermion] (4,0.5) --(5,0.5) ;
\draw[fermion] (3,0.5)--(5,1.5) ;
\draw[fermionloop] (5,1) circle (.5);
\draw[gauge] (4,0.5)  -- (5,-0.5);
\node at (5,1) {$^{3}S_{1}$};
\end{tikzpicture}
\caption{Feynman diagram for the $\protect\tau\to \protect\mu \overline{Q}%
Q[^{3}S_{1}]\protect\gamma$ decay.}
\label{tautomu3S1gamma}
\end{figure}
%%%%%
Neglecting the muon mass we obtain the following decay width%
\begin{equation}
\Gamma (\tau^{-} \rightarrow \mu^{-} ~\overline{Q}Q[^{3}S_{1}]\gamma )=\frac{\alpha
e_{Q}^{2}y^{2}|R(0)|^{2}}{32\pi ^{3}v^{2}}\frac{m_{\tau }^{3}M}{m_{H}^{4}}h\left(%
\frac{M^{2}}{m_{\tau }^{2}}\right),
\end{equation}%
where
\begin{equation}
h(x)=(1-x)^{3}+\frac{3}{2}x[(1-x)(3-x)+2\ln (x)].
\end{equation}%

\begin{figure}[ht]
\centering
\begin{tikzpicture}
\draw[gauge] (-3,0)--node[above]{$W$}(-1,0) ;
\draw[scalar] (-1,0)--node[above]{$H$}(1.5,-1) ;
\draw[gauge] (-1,0)--node[above]{$W$}(1,0.5) ;
\draw[fermion] (1,0.5) --(3,1) ;
\draw[fermion] (1,0.5)--(3,0) ;
\draw[fermionloop] (3.0,0.5) circle (.5);
\node at (3,0.5) {$\pi$};
\draw[fermion] (1.5,-1)  --node[above]{$\mu$}(3,-0.5) ;
\draw[fermion] (1.5,-1)  --node[below]{$\tau$}(3,-1.5) ;
\end{tikzpicture}
\caption{Feynman diagram for the $W\to\protect\tau\protect\mu \pi$ decay.}
\label{Wtotaumupi}
\end{figure}
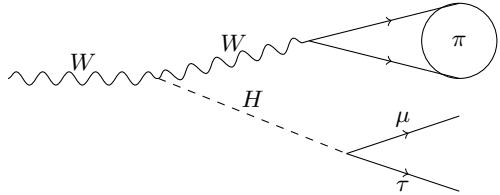

Finally, although the calculation does not require to use the quarkonium techniques, it is 
interesting  to estimate the effects of the $H\mu\tau$ coupling in LFV decay of gauge bosons. 
As a sample we calculate $W\to \mu \tau \pi $. The non-perturbative piece of this decay is 
related to the pion decay constant. This decay is induced by the diagram in Fig.(\ref{Wtotaumupi}).
%%%%%%
%%%%%

The amplitude for $W(Q,\protect\varepsilon )\rightarrow \protect\mu (p_{1})
\protect\tau (p_{2})\protect\pi (p_{3})$ is%
\begin{equation*}
\mathcal{M}=\frac{yg^{2}V_{ud}f_{\pi }}{2\sqrt{2}m_{W}}\frac{p_{3}\cdot
\varepsilon (Q)}{(p_{1}+p_{2})^{2}-m_{H}^{2}}\overline{u}(p_{2})v(p_{1}),
\end{equation*}%
where $g$ is the weak coupling constant. The resulting decay width is%
\begin{equation}
\Gamma (W^{-}\rightarrow \mu^{-} \tau^{+} \pi^{-} )=\frac{f_{\pi }^2 g^2 \lambda ^2 
V_{\text{ud}}^2}{73728   \pi ^3 m_{W}} f(a,b),
\label{GWmutaupi}
\end{equation}
where
\begin{eqnarray}
f(a,b) &=& \frac{1}{a^4}\left[ a^2 \left(b^2-1\right) \left(24 a^6-6 a^4
   \left(4 b^2+7\right)+a^2 \left(b^2+17\right)
   \left(2 b^2+1\right)-6 b^2\right)  \right.  \nonumber \\
  && - \left. 6\left(a^2-1\right)^2 \left(4 a^6-a^4 \left(6
   b^2+1\right)+2 a^2 b^4+b^4\right) \ln   \left(\frac{a^2-1}{a^2-b^2}\right)-12 b^4 \ln (b)\right],
\end{eqnarray}
with $a=\frac{m_{H}}{m_{W}}$, $b=\frac{m_{\tau}}{m_{W}}$ and we neglected the pion and muon masses.

%\section{Numerics}
The results for the studied decays depend on the color-singlet matrix elements $R(0)$ for
$^{3}S_{1}$ quarkonium and $R^{\prime}(0)$ for $^{3}P_{0}$ quarkonium. The same matrix elements
appear in the leptonic decay of the first and two photon decays of the latter. We use the 
available experimental results on these decays to extract the phenomenological value 
of the matrix elements. The only matrix element that cannot be calculated this way is 
$R^{\prime}(0)$ for the $^{3}P_{0}$ $\bar b b $ quarkonium $\chi_{b0}$. There is no available
experimental data on the $\chi_{b0} \to\gamma\gamma$ transition, but $R^{\prime}(0)$ has been 
calculated in several potential models summarized in Ref. \cite{Likhoded:2012hw} yielding 
$R^{\prime}(0)\approx 1 ~GeV^{5}$ and we will use this value in our calculations.

\begin{figure}[ht]
\centering
\begin{tikzpicture}
\draw[fermion] (-2,0.5) --(0,0) ;
\draw[fermion] (-2,-0.5)--(0,0) ;
\draw[fermionloop] (-2.0,0) circle (.5);
\draw[gauge] (0,0)--(2,0) ;
\node at (-2,0) {$^{3}S_{1}$};
\node at (1,0.2) {$\gamma$};
\node at(-1,0.7){$\frac{Q}{2}+q,s_{1}$};
\node at(-1,-0.7){$\frac{Q}{2}-q,s_{2}$};
\draw[fermion] (2,0)--node[above]{$l^{+}$}(3,0.5) ;
\draw[fermion] (2,0)--node[below]{$l^{-}$}(3,-0.5) ;
\end{tikzpicture}
\caption{Diagram for the $\overline{Q}Q[^{3}S_{1}]\to l^{+}l^{-} $ decay.}
\label{Vtoll}
\end{figure}
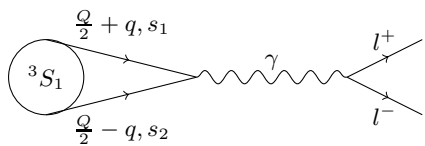

\begin{figure}[ht]
\centering
\begin{tikzpicture}
%First diagram
\draw[fermion] (-5,0.5) --(-3,0.5) ;
\draw[gauge] (-3,0.5)  -- node[above]{$k_1,\mu$}(-1,0.5);
\draw[fermion] (-3,0.5) --(-3,-0.5) ;
\draw[fermion] (-5,-0.5)--(-3,-0.5) ;
\draw[fermionloop] (-5.0,0) circle (.5);
\draw[gauge] (-3,-0.5)--(-1,-0.5) ;
\node at (-5,0) {$^{3}P_{0}$};
\node at (-2,-0.3) {$k_2,\nu$};
\node at(-4,0.7){$\frac{Q}{2}+q,s_{1}$};
\node at(-4,-0.7){$\frac{Q}{2}-q,s_{2}$};
%Second diagram
\draw[fermion] (1,0.5) --(3,0.5) ;
\draw[gauge] (3,-0.5)  -- node[above]{$k_1,\mu$}(5,-0.5);
\draw[fermion] (3,0.5) --(3,-0.5) ;
\draw[fermion] (1,-0.5)--(3,-0.5) ;
\draw[fermionloop] (1.0,0) circle (.5);
\draw[gauge] (3,0.5)--(5,0.5) ;
\node at (1.0,0) {$^{3}P_{0}$};
\node at (4,0.7) {$k_2,\nu$};
\node at(2,0.7){$\frac{Q}{2}+q,s_{1}$};
\node at(2,-0.7){$\frac{Q}{2}-q,s_{2}$};
\end{tikzpicture}
\caption{Feynman diagrams for the two photon decay of $^{3}P_{0}$ quarkonium.}
\label{Stogg}
\end{figure}
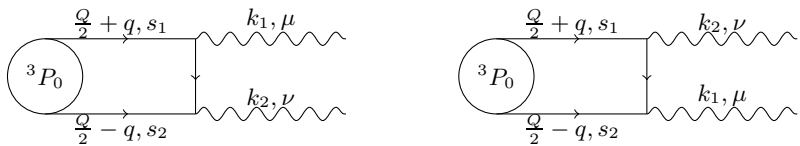

The leptonic decay of vector quarkonia is induced by the diagram in Fig.~(\ref{Vtoll}).
%%%%%%
%%%%%
The corresponding decay width is
\begin{equation}
\Gamma (\overline{Q}Q[^{3}S_{1}]\to l^{+}l^{-})=\frac{4\alpha^{2} e^{2}_{Q}|R(0)|^{2}}{M^2}
\end{equation}
where we neglected the lepton mass.
The two photon decay of $^{3}P_{0}$ quarkonium proceeds through the diagrams in Fig. (\ref{Stogg}).
%%%%%%
%%%%%
The decay width is
\begin{equation}
\Gamma(^{3}P_{0}\to \gamma\gamma)=\frac{432 \alpha^{2}e^{2}_{Q}|R^{\prime}(0)|^{2}}{M^{4}}
\end{equation}
In Table \ref{npme}. we give the results for the matrix elements of the different quarkonia. As a 
first approximation we use an $\bar{s}s$ configuration for the $f_{0}(980)$.
There is no available information on the total width of the $\chi_{b0}$, thus we report the decay 
width when we use $R^{\prime}_{\chi_b}(0)= 1~ GeV^{5}$ from quark model 
calculations \cite{Likhoded:2012hw}: 
\begin{equation}
\Gamma (\chi_{b0}\to \mu\tau)=5.5\times 10^{-17}~ GeV.
\end{equation}
 
The branching ratios of the remaining decays are calculated using the estimates for the 
non-perturbative matrix elements in Table  \ref{npme}. We list in Table \ref{BRsummary} 
the so obtained branching ratios.  Here, the branching 
ratios include a factor 2 to account for the two charge states where appropriate, e.g. 
$BR(\chi _{c0}\to \mu \tau)= BR(\chi _{c0}\to \mu^- \tau^+)+BR(\chi _{c0}\to \mu^+ \tau^-)$.

\begin{table}
\begin{tabular}{|c|c|c|c|}
\hline
Process & $\Gamma_{exp} (GeV)$ & $|R(0)|^{2}(GeV^{3})$ & $|R^{\prime}(0)|^{2} (GeV^{5})$ \\
\hline
\hline
$\Upsilon\to e^{+}e^{-}$ & $1.28\times 10^{-6}$ & $4.856$ & - \\
\hline
$J/\psi\to e^{+}e^{-}$ & $5.54\times 10^{-6}$ & $0.560$ & - \\
\hline
$\phi\to e^{+}e^{-}$ & $1.26\times 10^{-6}$ & $5.53 \times 10^{-2}$ & -\\
\hline
$\chi^{0}_{c}\to\gamma\gamma$& $2.34\times 10^{-6}$ & - & $3.10\times 10^{-2}$ \\
\hline
$f_{0}\to\gamma\gamma$ & $0.29\times 10^{-6}$ & - & $1.08\times 10^{-4}$ \\
\hline
\end{tabular}
\caption{Numerical values of the non-perturbative matrix elements extracted from the
leptonic and two photon decays of quarkonia.}
\label{npme}
\end{table}

\begin{table}
\begin{tabular}{|c|c|c|}
\hline
Process & Branching Ratio & Exp. bound \\
\hline
\hline
$\chi _{c0}\rightarrow \mu \tau $ & $1.5\times 10^{-17}$ & \\
\hline
$\Upsilon \rightarrow \mu \tau \gamma $ & $5.7\times 10^{-14}$ & \\
\hline
$J/\psi \rightarrow \mu \tau \gamma $ & $5.1\times 10^{-17}$ & \\
\hline
$\tau \rightarrow \mu f_{0}(980)$ & $8.4\times 10^{-12}$ & $< 3.4\times 10^{-8}$\\
\hline
$\tau \rightarrow \mu \phi \gamma $ & $1.7\times 10^{-14}$ & \\
\hline
%$\tau \rightarrow 3\mu $ & $2.3\times 10^{-12}$ & \\
%\hline
%$\tau \rightarrow \mu e^{+}e^{-}$ & $7.3\times 10^{-17}$ & \\
%\hline
$W\rightarrow \mu \tau \pi $ & $3.2\times 10^{-17}$ & \\
\hline
\end{tabular}%
\caption{Branching ratios for lepton flavor violation decays involving the $H%
\protect\mu\protect\tau$ coupling.}
\label{BRsummary}
\end{table}

In general these branching ratios are small. The most promising decay is the $\tau\to\mu f_{0}$. 
We recall that we assumed an $\bar{s}s$ configuration for this meson. The nature of the low 
lying scalar mesons is an old problem (see \cite{lightscalars} and references therein) and it would 
be desirable to have a closer approximation to the non-perturbative effects in this decay. 

\newpage
%%%%%%%%%%%%%%%%%%%%%%%%%%%%%%%%%%%%%%%%%%%%%%%%%%%%%%%%%%%%%%%%%%%%%%%%%%%%%%%%%%%%%%%%%%%%%%%%%%%%%%%%
\begin{acknowledgments}
We acknowledge financial support from CONACYT and SNI (M\'exico).D. D. is  grateful to Conacyt (M\'exico) S.N.I. and Conacyt project (CB-156618), DAIP
project (Guanajuato University) and PIFI (Secretaria de Educacion
Publica, M\'exico) for financial support.
\end{acknowledgments}

%%%%%%%%%%%%%%%%%%%%%%%%%%%%%%%%%%%%%%%%%%%%%%%%%%%%%%%%%%%%%%%%%%%%%%%%%%%%%%%%%%%%%%%%%%%%%%%%%%%%%%%

\end{document}